\documentclass[aps,prx,twocolumn,superscriptaddress,longbibliography]{revtex4-2}
\usepackage{graphicx}
\usepackage{subfig}

\begin{document}


\title{Sub-10 ps Minimum Ionizing Particle Detection with Geiger-Mode APDs
}


\author{Francesco Gramuglia}
\thanks{The first two authors contributed equally.}
\author{Emanuele Ripiccini}
\thanks{The first two authors contributed equally.}
\author{Carlo Alberto Fenoglio}
\author{Ming-Lo Wu}
\affiliation{École polytechnique fédérale de Lausanne (EPFL)}
\author{Lorenzo Paolozzi}
\affiliation{University of Geneva}
\affiliation{CERN}
\author{Claudio Bruschini}
\author{Edoardo Charbon}
\affiliation{École polytechnique fédérale de Lausanne (EPFL)}



\date{\today}

\begin{abstract}
Major advances in silicon pixel detectors, with outstanding timing performance, have recently attracted significant attention in the community. In this work we present and discuss the use of state-of-the-art Geiger-mode APDs, also known as single-photon avalanche diodes (SPADs), for the detection of minimum ionizing particles (MIPs) with best-in-class timing resolution. The SPADs were implemented in standard CMOS technology and integrated with on-chip quenching and recharge circuitry. Two devices in coincidence allowed to measure the time-of-flight of 180 GeV/c momentum pions with a coincidence time resolution of 22 ps FWHM (9.5 ps Gaussian sigma). Radiation hardness measurements, also presented here, highlight the suitability of this family of devices for a wide range of high energy physics (HEP) applications.
\end{abstract}


\maketitle

\section{Introduction}
\subsection{High timing resolution with silicon pixel detectors}
Silicon pixel detectors have been developed in high-energy physics applications to provide precise position measurements thanks to their compactness and high spatial granularity. Recent developments have been focused on sub-100 ps timing measurements of optical photons and direct detection of charged particles. 
\par When a particle passes through the detector, electron-hole pairs are generated. When these charges move in the depletion region, an induced current pulse is registered on one electrode. According to the Schockley-Ramo theorem \cite{Shockley_1938,Ramo_1939}, this current is proportional to the free charge $Q$, to the speed of the charge carriers $v$, and to the weighting field, which can be expressed, to a first approximation, as $\frac{1}{d}$, where $d$ is the thickness of the depletion region.
Hence, we can calculate the induced current as: 
\begin{equation}
  i=kQv\frac{1}{d},  
\end{equation}
where $k$ is a proportionality factor. The signal ends when all charges have been collected. Moreover, in case of a minimum ionizing particle (MIP) crossing a thin device, the charge $Q$ is proportional to $d$. We thus have:
\begin{equation}
i=kNdv\frac{1}{d}=kNv,
\end{equation}
where $N$ is the number of electron-hole pairs generated per unit length. This result shows that the initial value of the induced current is constant and does not depend on the thickness of the depletion region. When reading out this current signal on a load, behaving like an ideal transimpedance amplifier, we observe a sharp voltage pulse. The time-of-arrival (ToA) of the charge is usually determined by comparing the voltage pulse with a threshold. The uncertainty of the voltage pulse $\sigma_V$ is expressed as: 
\begin{equation}
\sigma_{V}=\sigma_{t}\frac{dV}{dt},
\label{eq:sigma_V}
\end{equation}
where $\sigma_t$ is the jitter of the voltage pulse. 
By inverting Eq. \ref{eq:sigma_V}, we find that:
\begin{equation}
\sigma_{t}=\frac{\sigma_{V}}{\frac{dV}{dt}}.
\label{eq:sigma_t}
\end{equation}
Eq. \ref{eq:sigma_t} shows that the signal fluctuation ($\sigma_{V}$) should be reduced to achieve a better timing jitter $\sigma_{t}$ and the slew rate $\frac{dV}{dt}$ should be maximized. In case of a sensor with an internal finite gain $G$ \cite{Cartiglia_2014,CARTIGLIA_2015,CARTIGLIA_2019}, the slew rate is proportional to $\frac{G}{d}$. This analysis suggests that thin sensors with large internal gain will in principle result in a better timing jitter. There is, however, another effect that can significantly affect the pulse shape while detecting MIPs: the charge collection noise. This phenomenon is caused by the variability of the profile of the deposited charge. As shown in \cite{Paolozzi_2015}, this effect introduces a timing jitter that is non-negligible at the 10 ps scale, and which increases with the detector thickness \cite{Jadhav_2021}. Various solutions have been proposed to reduce the contribution of this additional source of timing jitter, such as the detectors reported in \cite{CARTIGLIA_2019,LAI_2020,Anderlini_2020}. 
\par All the effects mentioned above call for extremely high intrinsic gain and slew rate together with thin structures. Thus, Geiger-mode silicon APDs, also known as single-photon avalanche diodes (SPADs), could represent promising candidates for substantial timing jitter reduction \cite{Ceccarelli_2020}. In SPADs, unlike APDs, the avalanche is a self-sustaining process, and the timing jitter contributions are more related to the avalanche growth dynamics. In particular, the timing jitter improves, decreasing the avalanche current value needed by the frontend electronics to detect a pulse \cite{Assanelli_2011}. A comprehensive theoretical study of timing performance in SPADs when used in MIP detection is presented in \cite{Riegler_2021}. First examples of such systems, detailed in \cite{DAscenzo_2014, vignetti_2017_thesis, Vignetti_2017}, were following the concept proposed by \cite{saveliev_2012}.\\
Although these potential advantages are promising, some problems usually affect Geiger-mode devices in the framework of MIP detection. Indeed, typical SPADs have a long dead time if no properly designed active quenching and recharge techniques are used \cite{Gallivanoni_2010}. Another issue is the presence of noise in the form of spurious pulses even in the dark. This noise, known as dark count rate (DCR), could limit the suitability of SPADs for the target application because of significant degradation of the measurement signal-to-noise ratio (SNR).
However, as shown by \cite{Ratti_2021, VIGNETTI_2018}, DCR can be drastically reduced by detecting MIPs with two SPADs operated in coincidence. Building on these elements, we present MIP time-of-flight (ToF) measurements resulting in unprecedented timing precision. 

\subsection{SPAD Detector and System-on-Board}
 \begin{figure}[t]
 \includegraphics[width=\columnwidth]{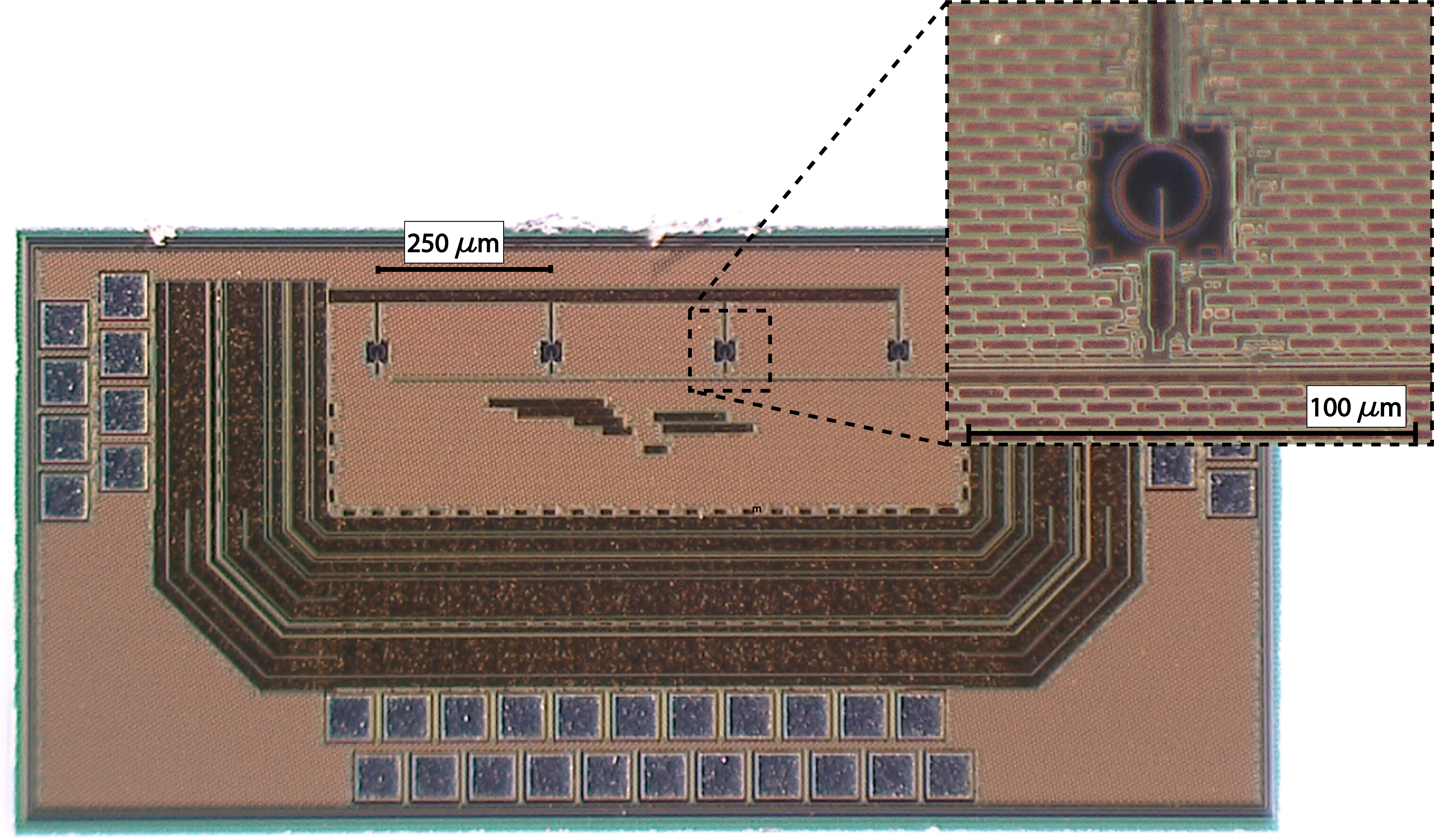}%
 \caption{Micrograph of the implemented chip embedding 25 $\mu m$ diameter SPADs with integrated pixel circuit \cite{Gramuglia_2021}.}
 \label{fig:micrograph}
 \end{figure}
 \begin{figure}[t]
 \includegraphics[width=\columnwidth]{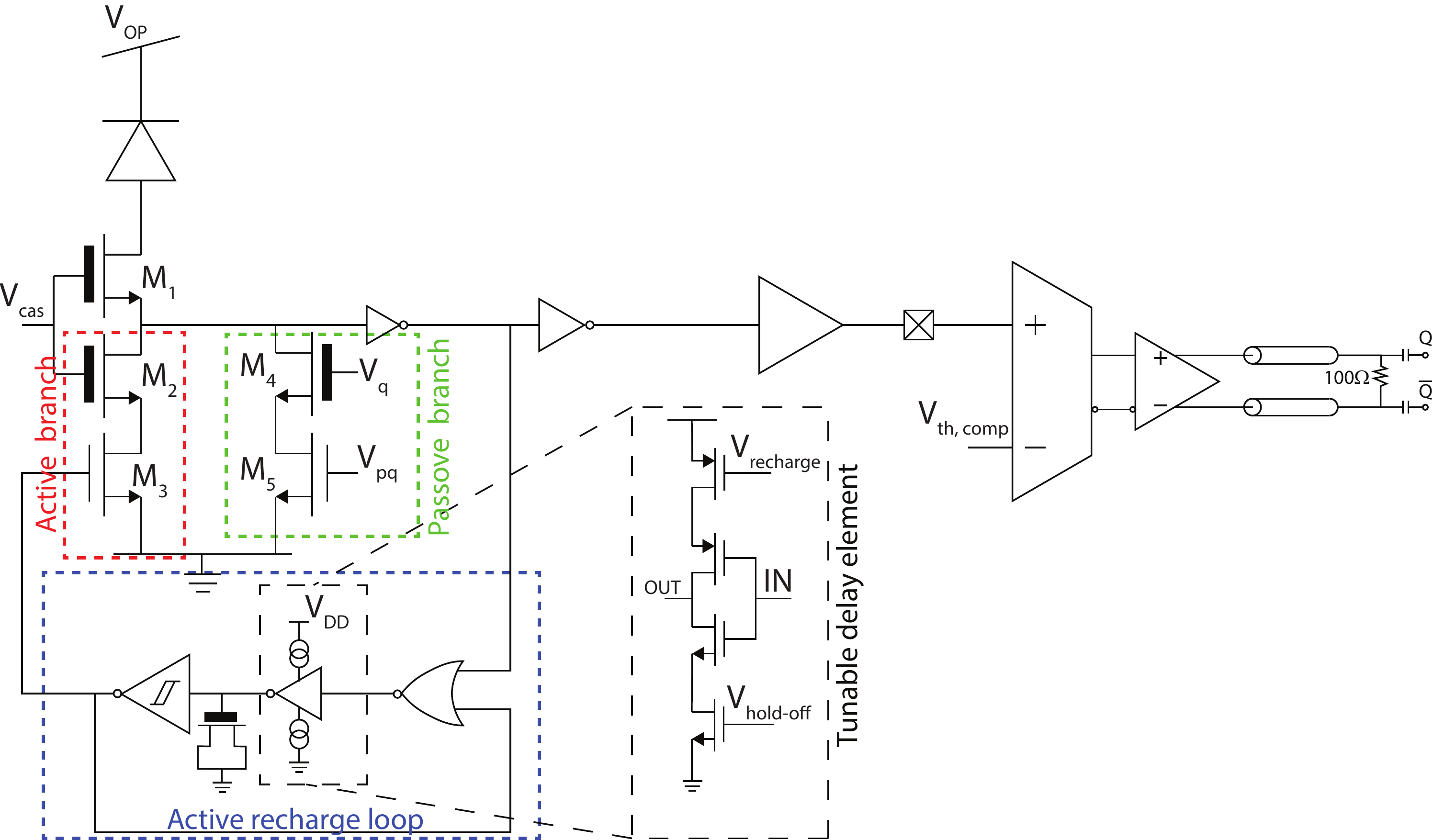}%
 \caption{Device schematic. The chip output is connected to a fast SiGe comparator to drive a 50 $\Omega$ differential output line.}
 \label{fig:schematic}
 \end{figure}
The detector system used in this work relies on the SPAD-based sensor presented in \cite{Gramuglia_2021}. FIG.~\ref{fig:micrograph} shows a micrograph of the sensor used in this study. The sensor integrates four independent SPAD pixels with a diameter of 25 $\mu$m. A dedicated on-chip front-end circuitry, shown in FIG. \ref{fig:schematic}, is implemented in close proximity to each SPAD. The circuit is designed to enable a tunable dead time, as short as 3 ns, supporting very high count rates while still maintaining very low afterpulsing \cite{Gramuglia_2021}.
\par In this work, we implemented a complete and optimized system-on-board to further improve performance. The resulting system comprises a motherboard, where all needed voltage levels are derived from a single 5 V power supply. A power management unit was designed to filter most of the electronic noise and to provide low-noise and stable voltage levels to the detector and its front-end circuits. Moreover, the full integration of a system-on-board reduces the noise picked up in cables and power cords that can act as antennas. Indeed, reducing the noise in the system is essential when the target timing precision approaches 10 ps  \cite{Cartiglia_2014}. A full system control is achieved with a serial bus interface that allows the tuning of the device operating point from an host computer. \\
The output of the chip is connected to fast SiGe comparators (Analog Device ADCMP572) that drive 50 $\Omega$ lines (FIG. \ref{fig:schematic}). This solution reduces the capacitive load at the chip's output (high impedance node) and helps propagate the signal through a high-frequency cable to the timestamping electronics. In addition, the use of these comparators makes it possible to achieve high signal slew rate ($\geq$ 1.6 V/ns).

\section{Optical Device Characterization}
\label{ch:setup}
\begin{figure}
 \includegraphics[width=\columnwidth]{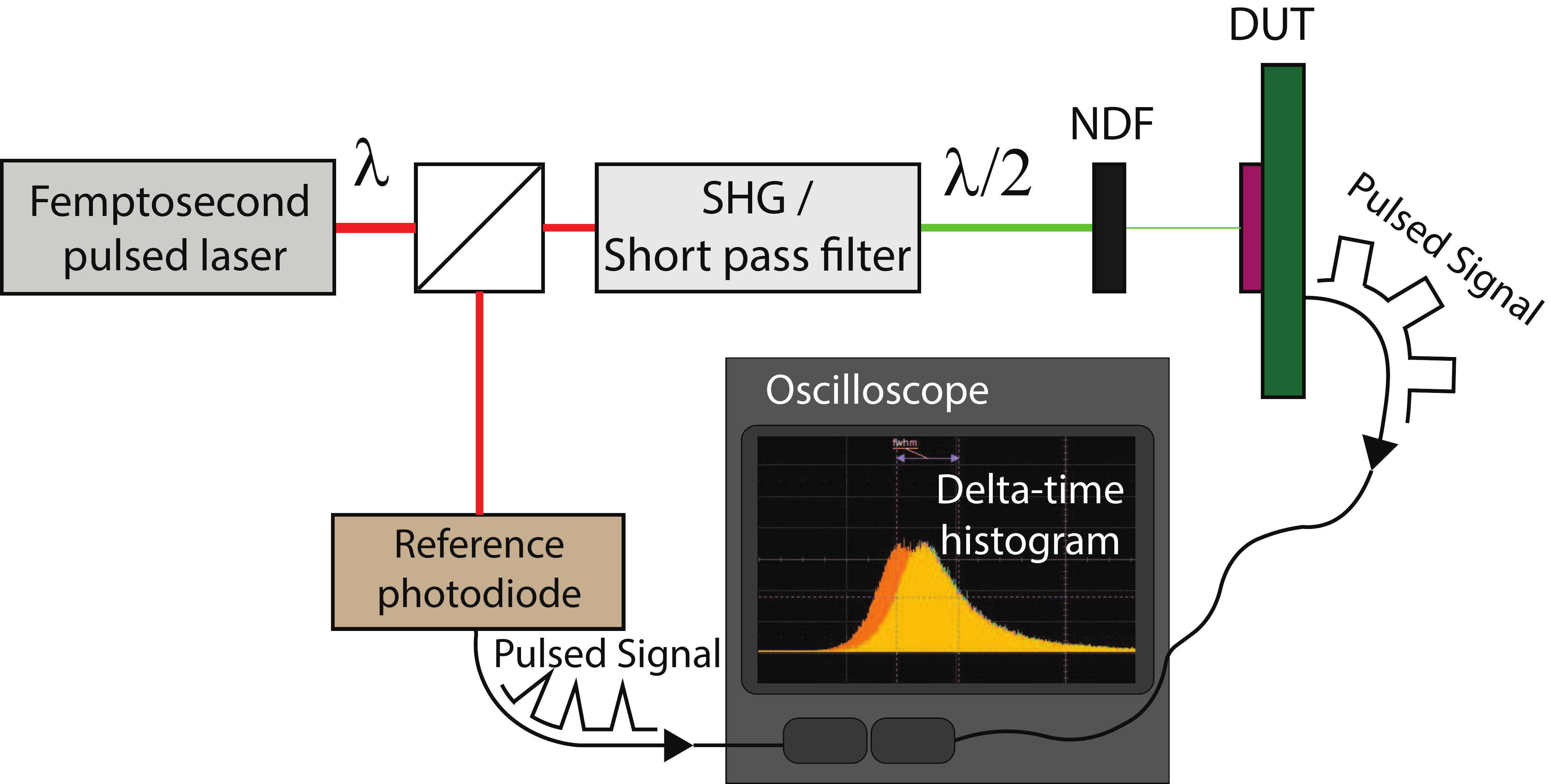}%
 \caption{Experimental setup for device characterization using femtosecond pulsed lasers.}
 \label{fig:light_setup}
 \end{figure}

In order to analyze the performance of our system, we started from an optical characterization using the setup in FIG. \ref{fig:light_setup} \cite{Gramuglia_2021,Gramuglia_2021_4}. The test bench is composed of a femtosecond pulsed laser used as the controlled light source, making the pulse length contribution to the measured timing jitter negligible. The laser beam is split into two branches. One branch is captured by a fast photodiode used as an optical, rather than electrical, reference to ensure that the dominant jitter is that of the device under test (DUT). The other branch passes through a second harmonic generation (SHG) stage to generate a light pulse in the visible range (i.e., within our device sensitivity spectrum). The latter is then attenuated, employing a neutral density filter (NDF) to reach a single-photon regime, and sent to the DUT. Finally, the output signals of the photodiode and the DUT are connected to an oscilloscope to build a time-difference histogram, representing the instrument response function (IRF). An asymmetric curve characterizes the typical SPAD IRF with the main peak, generally modeled with a Gaussian profile and an exponential tail  \cite{Lacaita_1993,Gulinatti_2011_2,RIPAMONTI_1985}. 
\par
\begin{figure}
  \subfloat[]{\includegraphics[width=\columnwidth]{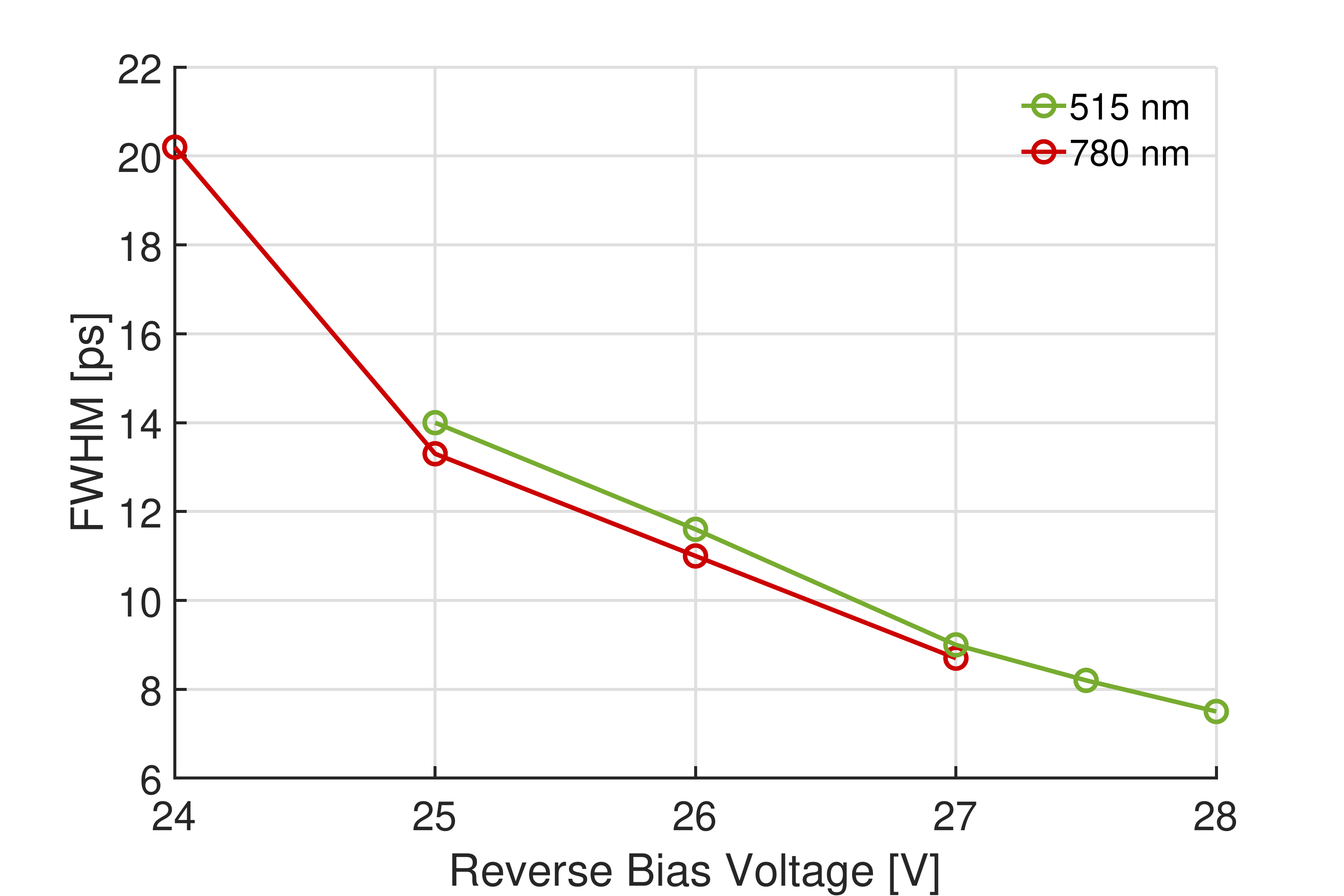}}\\
  \subfloat[]{\includegraphics[width=\columnwidth]{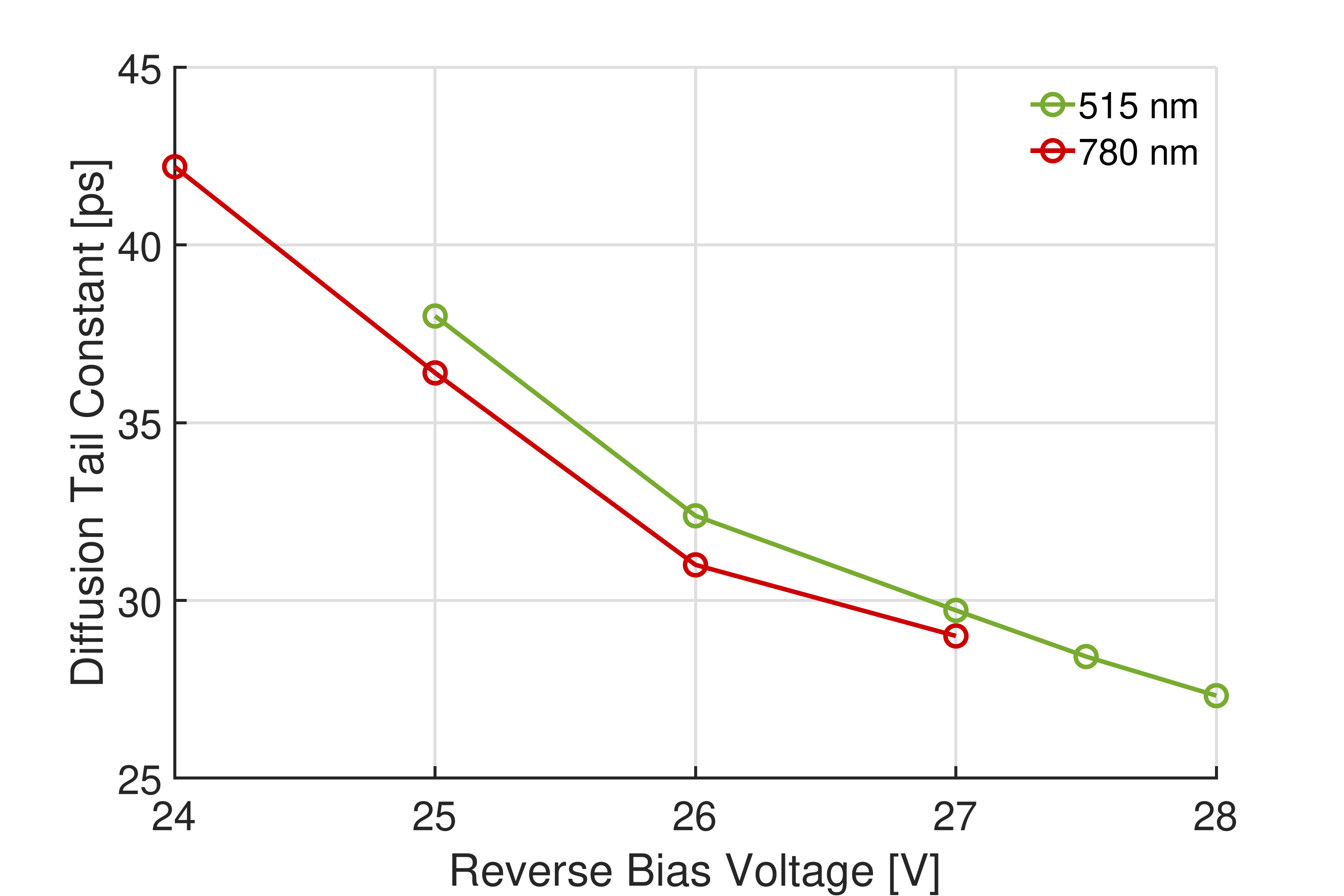}}%
 \caption{$(a)$ FWHM single-photon timing resolution obtained with the designed system. $(b)$ Exponential tail time constant measured at various excess bias voltages. For these experiments two wavelengths were used: 515 nm and 780 nm. The results show a jitter of $\sim$7.5 ps FWHM for green and $\sim$27 ps FWHM for red light at $\sim$6.5 V of excess bias voltage.}
 \label{fig:light_result}
 \end{figure}

 The experiment has been repeated for several excess bias voltages and two wavelengths, and the results are shown in FIG. \ref{fig:light_result}. The timing results are expressed as FWHM of the IRF. The timing precision shows a dependency on the bias point, and it improves when increasing the applied voltage, as expected. With this improved system, we reached a timing jitter of 7.5 ps at a reverse bias voltage of 28 V, corresponding to an excess bias ($V_{ex}$) of about 6.5 V. Moreover, we report in FIG. \ref{fig:light_result} $(b)$ the decay time constant of the exponential tail again as a function of reverse bias voltage. These results show an improvement of almost 40\% with respect to what reported in \cite{Gramuglia_2021}, where the output was directly taken from the packaged die with high impedance 4 GHz active probes.
The mean free path of a MIP in silicon is on the order of hundreds of nanometers. Therefore, considering the thin structure of the proposed SPADs, we expect only a small amount of generated charge inside the sensitive volume of the device. Moreover, Geiger-mode operation and a prompt avalanche detection guarantee no difference in the output signal amplitude. For this reason, this experiment gives us a reasonable estimation of the expected system performance when detecting a MIP.
\section{ToF measurements for Minimum Ionizing Particles}
 \begin{figure}[t]
 \includegraphics[width=\columnwidth]{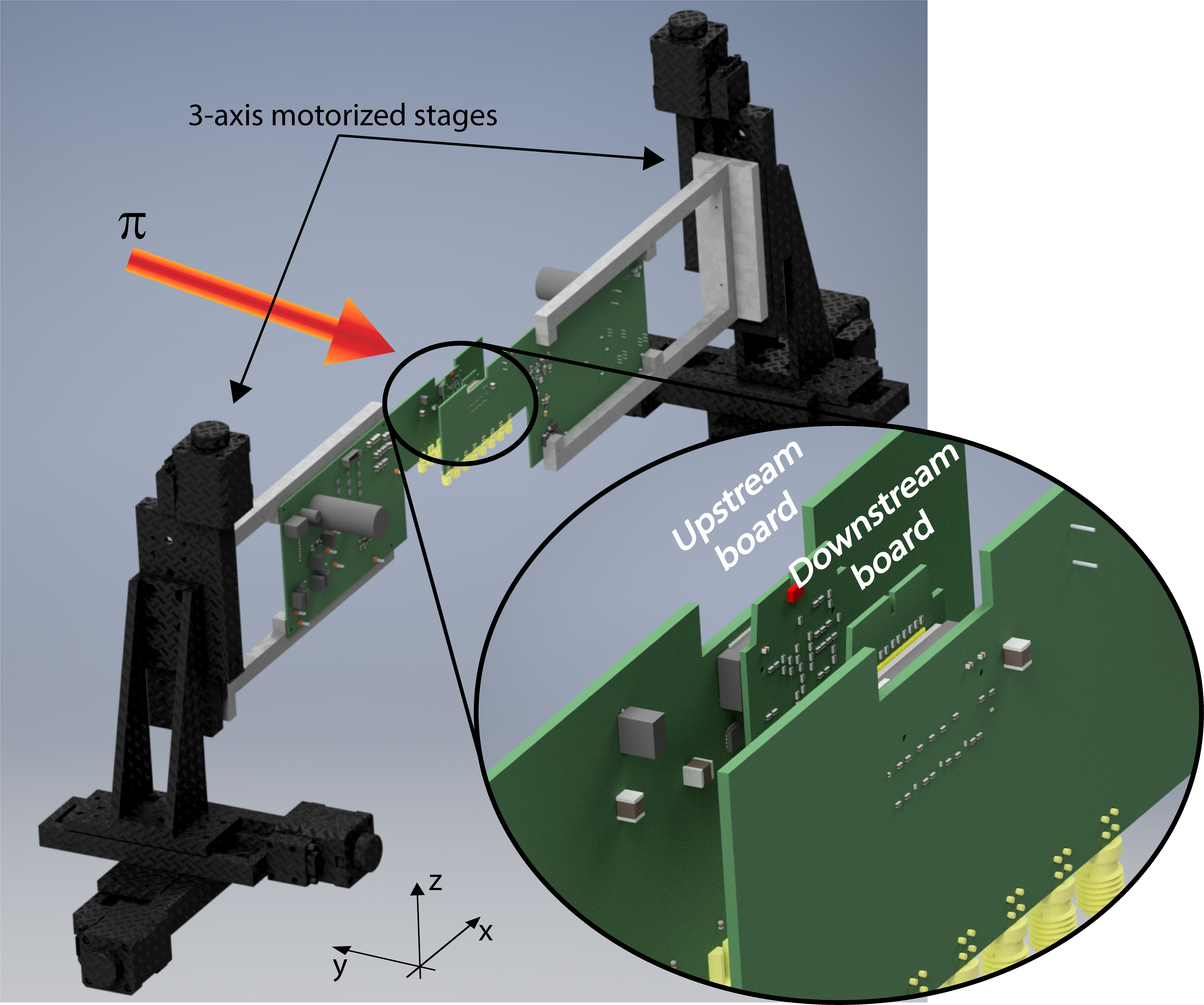}%
 \caption{Experimental MIP coincidence measurement setup.}
 \label{fig:setup}
 \end{figure}
 \begin{figure}[!ht]
 \subfloat[]{\includegraphics[width=\columnwidth]{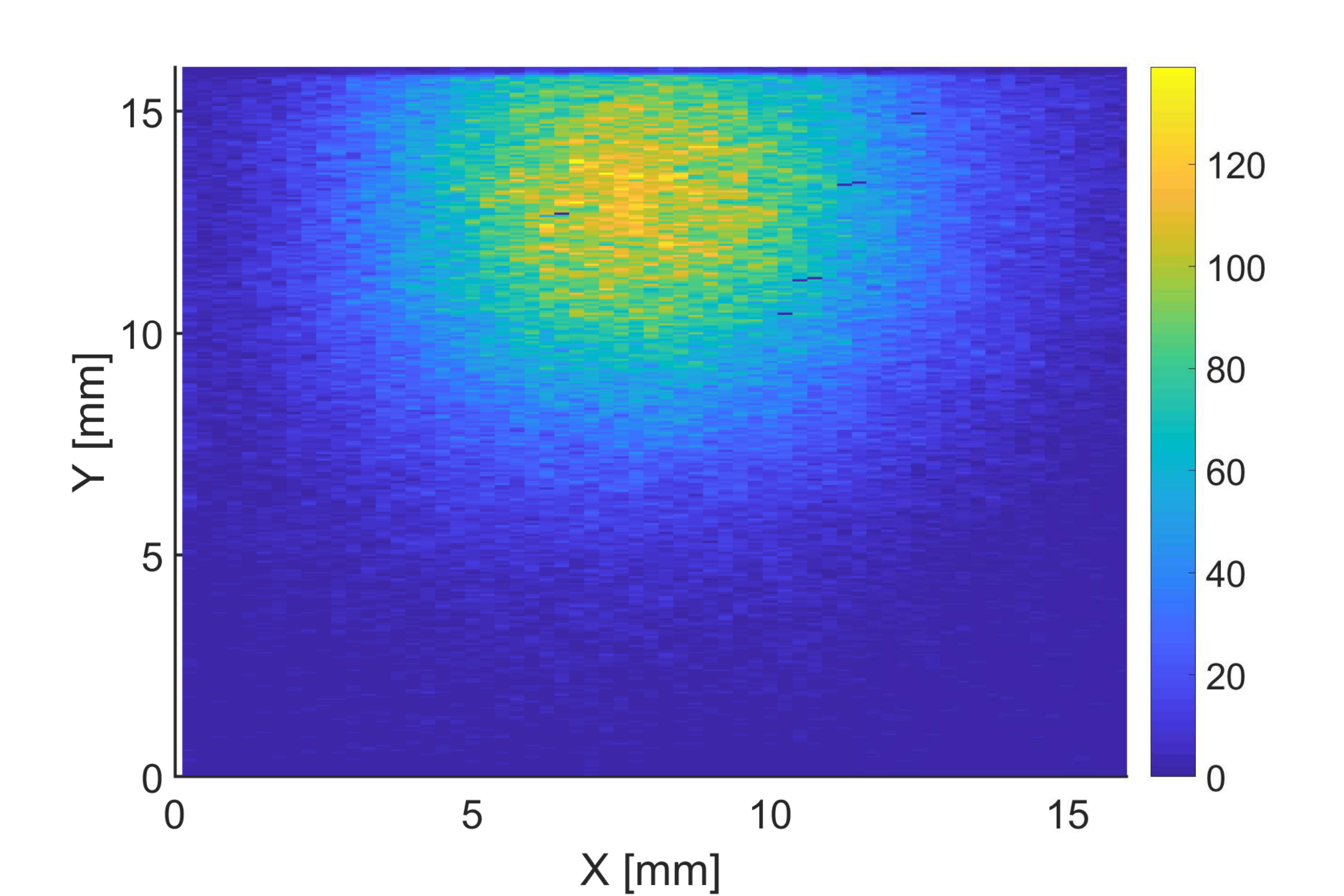}}\\
 \caption{MIP beam profile acquired with the HVCMOS telescope described in \cite{Terzo_2019}.}
 \label{fig:beamprofile}
 \end{figure}
 \begin{figure}[!ht]
 \subfloat[]{\includegraphics[width=\columnwidth]{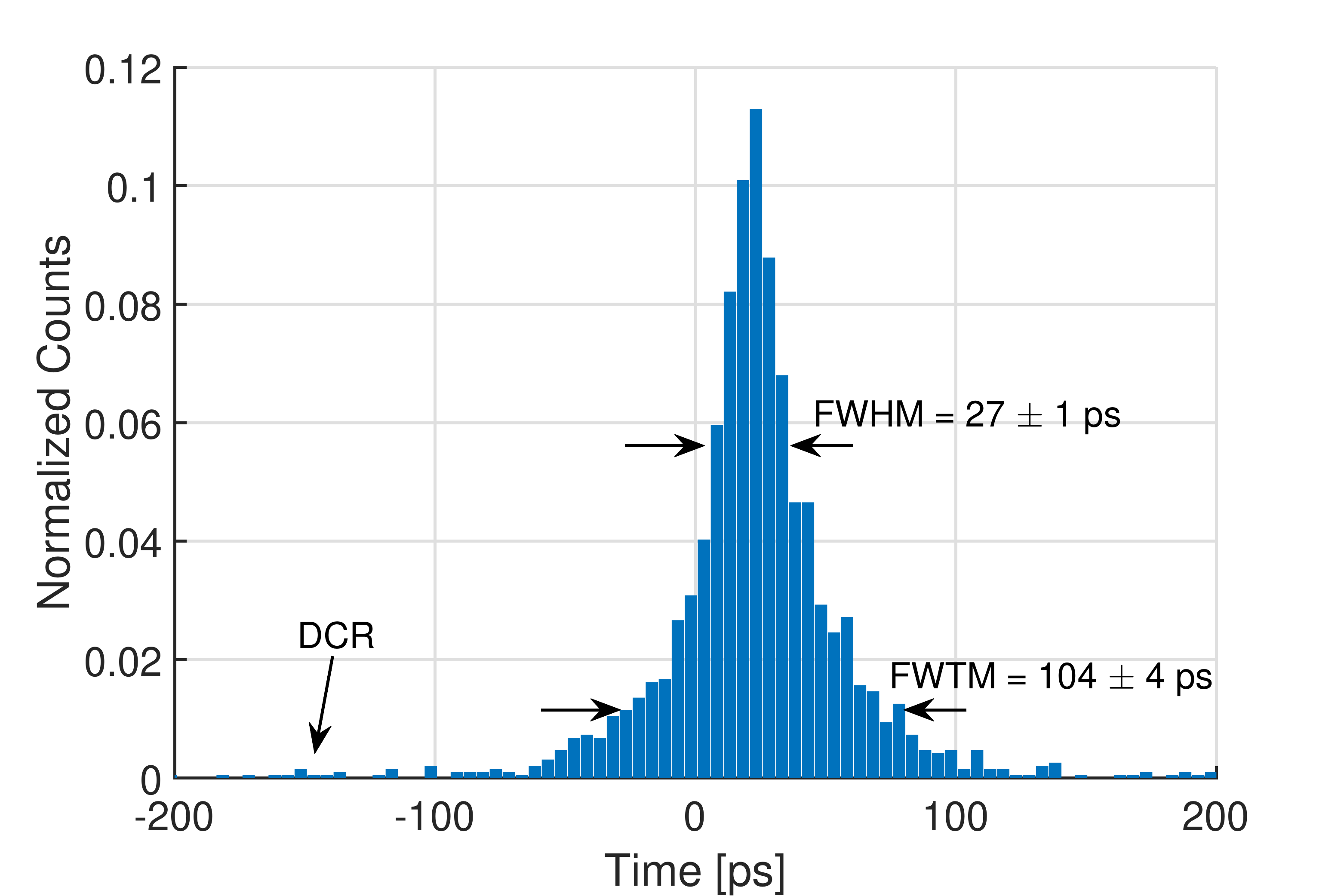}}\\
  \subfloat[]{\includegraphics[width=\columnwidth]{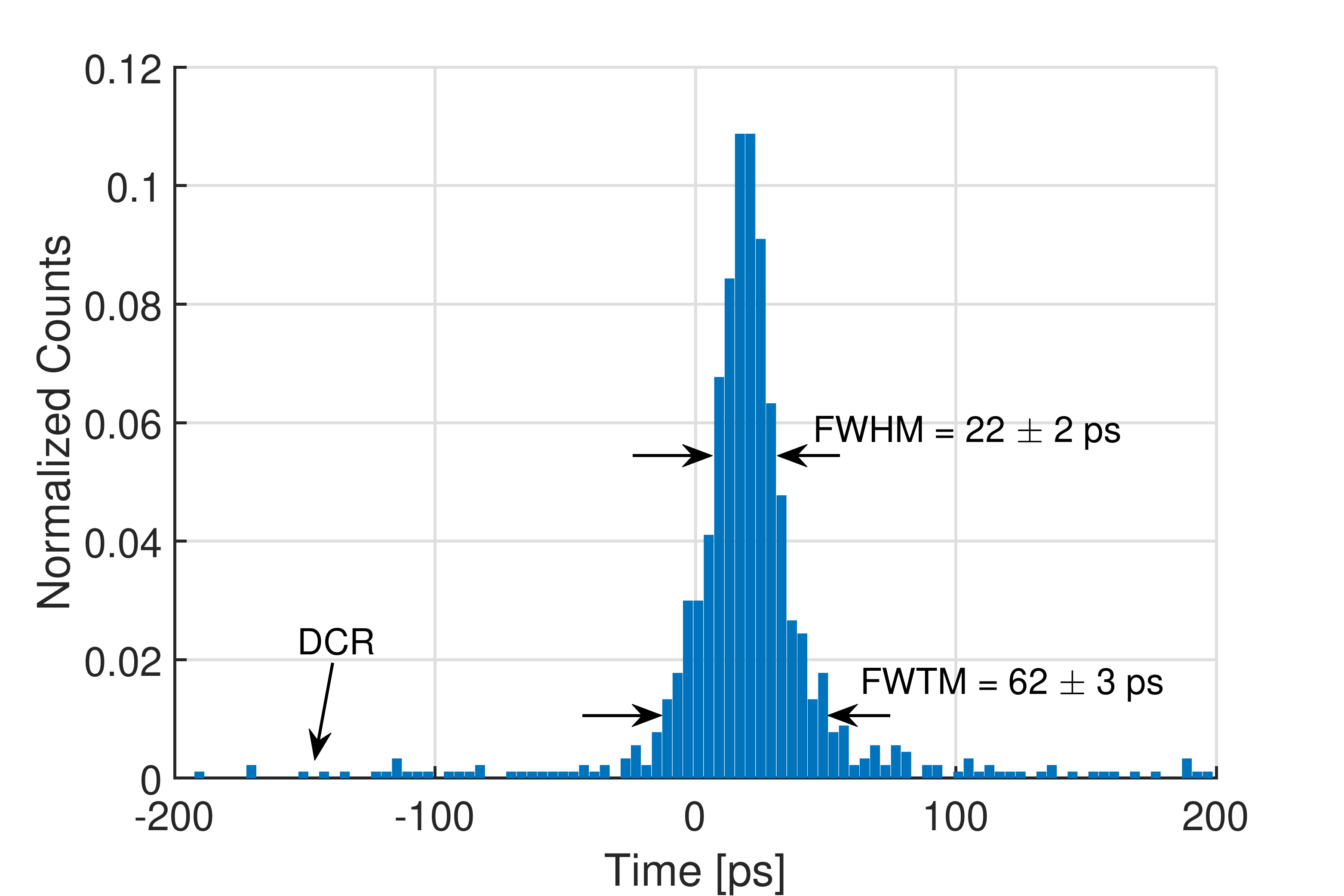}}
 \caption{Normalized distribution of the MIP time-of-flight between two SPADs in coincidence. $(a)$ at 24 V and $(b)$ at 27 V bias voltage.}
 \label{fig:TOF}
 \end{figure}
The setup used for the ToF measurement of MIPs is shown in FIG. \ref{fig:setup}. It consists of two systems-on-board (see Section \ref{ch:setup}), both mounted on motorized linear stages with sub-micrometer positioning resolution to allow a proper detector alignment and to guarantee the acquisition of coincidence measurements. We installed the setup on the H8 beamline in the CERN North Area. This beamline delivers 180 GeV/c momentum pions produced on a graphite target by the interaction of protons accelerated by the Super Proton Synchrotron (SPS).
\par The beam profile measurement is reported in Fig. \ref{fig:beamprofile}. A good alignment between the two detectors has been achieved thanks to an HVCMOS telescope \cite{Terzo_2019}. 
The two detectors were positioned at the center of the beam, where the intensity is the highest. Coincidence events were acquired for two bias voltages, 24 V and 27 V, corresponding to approximately 2.5 V and 5.5 V $V_{ex}$, respectively. 
FIG. \ref{fig:TOF} shows the ToF distributions for both $V_{ex}$. The MIP measurement results are summarized in TABLE \ref{tab:analysis_table}.
 \begin{table}[t]
    \centering
    \scriptsize
 \begin{center}
\begin{tabular}{ c|c|c|c|c} 
 \hline
 Bias (V) & FWHM (ps) & FWTM (ps) & $\sigma$ (ps) & $\sigma_{single}$ (ps) \\ 
 \hline
 24 & 27 $\pm$ 1 & 104 $\pm$ 4 & 11.5 $\pm$ 0.4 & 8.1 $\pm$ 0.3\\ 
 27 & 22 $\pm$ 2 & 62 $\pm$ 3 & 9.4 $\pm$ 0.7 & 6.6 $\pm$ 0.5 \\
 \hline
\end{tabular}
\end{center}
 \caption{Summary of the MIP detection measurement results. The Gaussian sigma has been obtained by dividing the FWHM by $2\sqrt{2ln(2)}$. Assuming that the response is the same for both SPADs, the $\sigma_{single}$ values have been obtained by dividing the $\sigma$ values by $\sqrt{2}$. The errors have been evaluated using statistical error propagation.}
    \label{tab:analysis_table}
\end{table}

\section{Radiation Hardness}
\begin{figure}[t]
 \includegraphics[width=\columnwidth]{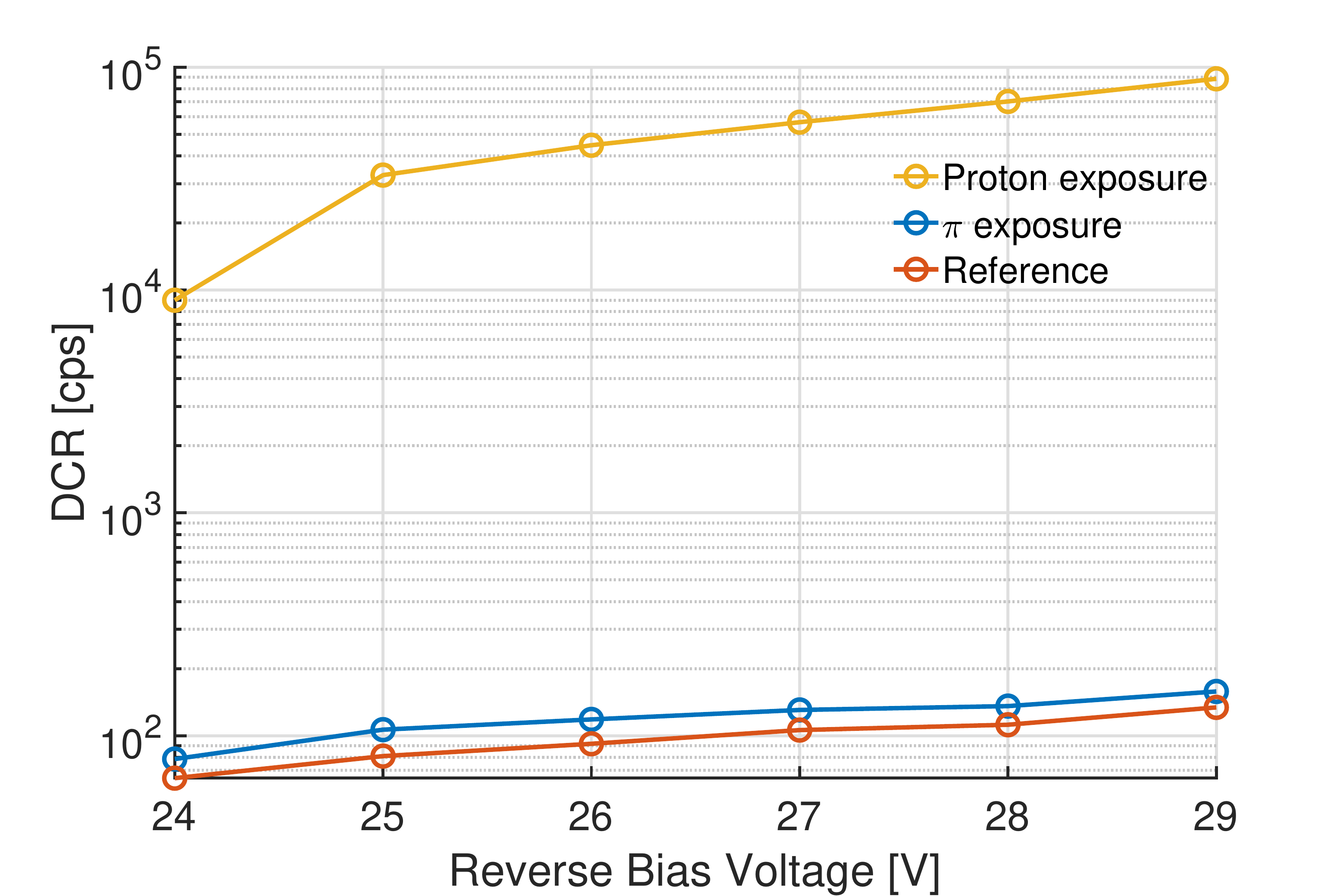}%
 \caption{Radiation hardness experimental results. DCR trends over various excess bias voltages are reported for the same device. The reference with no irradiation (\emph{orange}) is compared with the noise level after 100 MeV proton irradiation (\emph{yellow}), and with the samples after 1 week of 180 GeV/c pions irradiation.}
 \label{fig:radhard}
 \end{figure}

The radiation hardness of the SPAD detectors was characterized by using protons which induce both ionizing damage and displacement damage. The detectors were irradiated using the Proton Irradiation Facility (PIF) at the Paul Scherrer Institute (PSI, Villigen, Switzerland). We used a 100 MeV mono-energetic beam with a fluence of $1\times10^{8}$ protons per second to reach a 300 TeV/g displacement damage dose (DDD) and 9.4 krad total ionizing dose (TID). The DCR difference was measured two weeks after the exposure with the aforementioned setup. The DCR comparison before and after the exposure is reported in FIG.\ref{fig:radhard}.
The characterization of the radiation hardness shows that the SPAD detector can maintain its functionality under the given radiation dose. The SPADs are not saturated by the DCR induced by the radiation damage thanks to their short dead time and high count rate. Moreover, as the ToF measurement is based on coincidence, even detectors reaching a DCR value of $1\times10^{5}$ counts per second will not affect their particle detection performance, as also shown in \cite{Ratti_2021, VIGNETTI_2018}. No other degradation of the device performance was observed.
\section{Discussion}
In this work we showed how Geiger-mode devices (i.e., SPADs) can detect MIPs with a sub-10 ps timing precision. This result paves the way to the implementation of future high timing resolution particle trackers based on this kind of detector. Moreover, the radiation hardness of the device was explored up to a DDD of 300 TeV/g. After exposure, we observed an increment of the DCR by about 3 orders of magnitude. Nevertheless, this increment does not affect the timing performance of the device. In fact, the number of accidental coincidences due to DCR is strongly suppressed by the logic AND between the two SPADs used for the ToF measurement \cite{Ratti_2021}.
\par When analyzing FIG. \ref{fig:TOF} and the results in TABLE \ref{tab:analysis_table}, we notice a dependency of the performance on the applied bias voltage. In particular, we can see a lower FWTM and an improvement in the FWHM for the ToF distribution when increasing the bias point from 24 V to 27 V. The higher field improves the avalanche buildup and lateral spread time dispersion \cite{Spinelli_1997,Tan_2007,Ingargiola_2009}. Moreover, a higher bias voltage enlarges the drift region and increases the electric field inside it. This effect reduces the size of the neutral region and helps minimize the statistical spread of the diffusion and transit time needed by the primary charge carriers to reach the multiplication region \cite{Ceccarelli_2020}.
\\Both distributions show a negligible flat background coming from random dark count coincidences. This is indicative of the efficacy of the noise filtering provided by measurements in coincidence.
\section{Acknowledgments}

We would like to thank the FASER group of the DPNC of the University of Geneva, led by Prof. Iacobucci, for the help and the support during the beamtime. This research was supported, in part, by the Swiss National Science Foundation under grant 200021-169465 and Sinergia CRSII5-177165.

\bibliography{./Testbeam.bib}

\end{document}